\documentclass[epj,final]{svjour}

\usepackage{bm}
\usepackage{cancel}
\usepackage{graphicx}
\usepackage{psfrag}
\usepackage{subfigure}

\begin{document}

\title{
	Relating the dynamics of road traffic in a stochastic cellular automaton to a 
	macroscopic first-order model
}

\titlerunning{
	Relating the dynamics of road traffic in a stochastic CA to a macroscopic 
	first-order model
}

\author{
  Sven Maerivoet\inst{1} \and Steven Logghe\inst{2} \and Ben Immers\inst{3} \and Bart De Moor\inst{1}
}

\authorrunning{
  Sven Maerivoet et al.
}

\institute{
  Department of Electrical Engineering ESAT-SCD (SISTA)\\
  Katholieke Universiteit Leuven\\
  Kasteelpark Arenberg 10, 3001 Leuven, Belgium\\
  Phone: +32 (0)16 32 17 09 Fax: +32 (0)16 32 19 70\\
  URL: \texttt{http://www.esat.kuleuven.ac.be/scd}\\
  \email{sven.maerivoet@esat.kuleuven.ac.be}
  \and
  Transport \& Mobility Leuven\\
  Vital Decosterstraat 67A bus 0001, 3000 Leuven, Belgium\\
  Phone: +32 (0)16 31 77 30 Fax: +32 (0)16 31 77 39\\
  URL: \texttt{http://www.tmleuven.be}\\
  \email{steven@tmleuven.be}
  \and
  Department of Civil Engineering
  Katholieke Universiteit Leuven\\
  Kasteelpark Arenberg 40, 3001 Leuven, Belgium\\
  Phone: +32 (0)16 32 16 69 Fax: +32 (0)16 32 19 76\\
  URL: \texttt{http://www.esat.kuleuven.ac.be/scd}\\
  \email{ben.immers@bwk.kuleuven.be}
}

\date{Received: June 28, 2005 / Revised version: \today}

\abstract{
	In this paper, we describe a relation between a microscopic traffic cellular 
	automaton (TCA) model (i.e., the stochastic TCA model of Nagel and 
	Schreckenberg) and the macroscopic first-order hydrodynamic model of 
	Lighthill, Whitham, and Richards (LWR). The innovative aspect of our approach, 
	is that we explicitly derive the LWR's fundamental diagram directly from the 
	STCA's rule set, by assuming a stationarity condition that converts the STCA's 
	rules into a set of linear inequalities. In turn, these constraints define the 
	shape of the fundamental diagram, which is then specified to the LWR model. 
	Application of our methodology to a simulation case study, allows us to 
	compare the tempo-spatial behavior of both models. Our results indicate that, 
	in the presence of noise, the capacity flows in the derived fundamental 
	diagram are overestimations of those of the STCA model. Directly specifying 
	the STCA's capacity flows to the LWR fundamental diagram, effectively remedies 
	most of the mismatches between both approaches. Our methodology sees the STCA 
	complementary to the LWR model and vice versa, so the results can be of great 
	assistance when interpreting the traffic dynamics in both models. Especially 
	appealing, is the fact that the STCA can visualize the higher-order 
	characteristics of traffic stream dynamics, e.g., the fans of rarefaction 
	waves.
}

\maketitle


\newcommand{\psfragstyle}[1]{\small{#1}}

\newcommand{\figref}[1]{Fig.~\ref{#1}}

\newcommand{\tableref}[1]{Table~\ref{#1}}

\newcommand{\eqref}[1]{(\ref{#1})}

\newlength{\tempfboxsep}
\setlength{\tempfboxsep}{\fboxsep}

\newcommand{\zero}{\ensuremath{\mbox{0}}}
\newcommand{\one}{\ensuremath{\mbox{1}}}
\newcommand{\two}{\ensuremath{\mbox{2}}}
\newcommand{\three}{\ensuremath{\mbox{3}}}
\newcommand{\four}{\ensuremath{\mbox{4}}}
\newcommand{\five}{\ensuremath{\mbox{5}}}
\newcommand{\ten}{\ensuremath{\mbox{10}}}

\newlength{\figurewidth}
\setlength{\figurewidth}{8.6cm}

\newlength{\figurehalfwidth}
\setlength{\figurehalfwidth}{0.48\figurewidth}


	\section{Introduction}

Considering the existing relations between the stochastic TCA model of Nagel and 
Schreckenberg \cite{NAGEL:92,NAGEL:95c} and the macroscopic first-order model of 
Lighthill, Whitham, and Richards (LWR) \cite{LIGHTHILL:55,RICHARDS:56}, from 
traffic flow theory, there are already numerous links between both modeling 
approaches. An example is the so-called totally asymmetric simple exclusion 
process (TASEP) \cite{DERRIDA:92}, which corresponds to the LWR model with a 
noisy and diffusive conservation law if a random sequential update is assumed 
\cite{NAGEL:95c,NAGEL:96}. Another example is the STCA which can be approximated 
by a so-called mean field theory (MFT), and its successive refinements, such as 
the car-oriented mean-field theory (COMF), and the recently developed 
site-oriented cluster-theoretic approach 
\cite{SCHRECKENBERG:95,SCHADSCHNEIDER:97,SCHADSCHNEIDER:98,SANTEN:99,SCHADSCHNEIDER:99b,CHOWDHURY:00,SCHADSCHNEIDER:02}.
A more elaborate discussion on these analytical treatments can be found in 
\cite{MAERIVOET:05}. In summary, we can say that there already exist several 
methods for bridging both the microscopic STCA and the macroscopic LWR model 
(note that we do not consider the class of hybrid models, as we are only 
interested in direct analogies between both microscopic and macroscopic models, 
and not in pure combinations of these model classes).

In this paper, we reconsider the STCA and LWR models, but we take a different 
approach at studying their relation: we consider a TCA model as a particle-based 
discretization scheme for macroscopic traffic flow models. It is from this 
latter point of view that our work addresses the common structure between both 
models. Our main goal is therefore to provide a means for implicitly 
incorporating the STCA's stochasticity into the LWR model, which is in fact 
deterministic in nature \cite{MAERIVOET:03g}. Note that we use the term 
\emph{implicit} to denote the fact that the STCA's stochasticity is not 
introduced in the equations by means of \emph{explicit} noise terms. Rather, our 
methodology implies that the stochasticity is introduced through the shape of 
the LWR's fundamental diagram.\\

This paper is organized as follows: in section \ref{sec:Recapitulation}, we 
briefly recapitulate both the approaches taken by traffic cellular automata 
models (with the STCA in particular), and the macroscopic first-order 
hydrodynamic LWR model. In section \ref{sec:InitialMethodology}, we then 
consider a methodology for implicitly incorporating the STCA's stochasticity 
into the LWR's triangular fundamental diagram. Continuing, we apply this 
technique to a small case study in section \ref{sec:ApplicationToACaseStudy}, 
which points us to some discrepancies between both modeling approaches. 
Highlighting some of the resulting artifacts, and investigating the main reason 
for the difference in behavior, we move on to section 
\ref{sec:AlternateDerivation} where we present an alternate derivation of the 
fundamental diagrams. Finally, the paper concludes with section 
\ref{sec:Conclusions}, stating a summary of our findings.

	\section{Recapitulating the STCA and LWR models}
	\label{sec:Recapitulation}

With respect to the modeling of traffic flows, there are largely two model 
classes possible, i.e., the microscopic and macroscopic approach, respectively. 
In the former class, interactions between vehicles in a traffic stream are 
explicitly modeled, giving rise to car-following and lane-changing submodels. 
With respect to the latter class, traffic streams are mostly treated as inviscid 
but compressible fluids. In this section, we briefly recapitulate a special 
class of microscopic models, i.e., traffic cellular automata (TCA) models. We 
then describe a stochastic TCA model, called the stochastic traffic cellular 
automaton (STCA) of Nagel and Schreckenberg, after which we conclude with an 
overview of the most prominent features of the macroscopic first-order 
hydrodynamic model of Lighthill, Whit\-ham, and Richards (LWR).

		\subsection{Traffic cellular automata (TCA) models}
		\label{sec:TCAModels}

\begin{figure}[!ht]
  \centering
  \psfrag{t}[][]{\footnotesize{$t$}}
  \psfrag{t+1}[][]{\footnotesize{$t + 1$}}
  \psfrag{i}[][]{\footnotesize{$i$}}
  \psfrag{j}[][]{\footnotesize{$j$}}
  \psfrag{deltat}[][]{\footnotesize{$\Delta T$}}
  \psfrag{deltax}[][]{\footnotesize{$\Delta X$}}
  \psfrag{gsi}[][]{\footnotesize{$g_{s_{i}}$}}
  \includegraphics[width=\figurewidth]{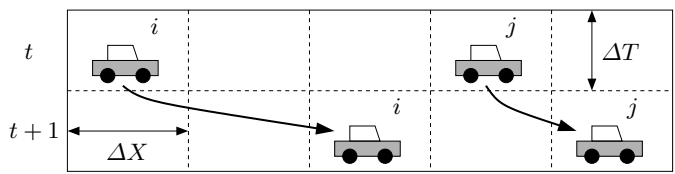}
  \caption{
    Schematic diagram of the operation of a single-lane traffic cellular 
    automaton (TCA); here, the time axis is oriented downwards, the space axis 
    extends to the right. The TCA's configuration is shown for two consecutive 
    time steps $t$ and $t + 1$, during which two vehicles $i$ and $j$ propagate 
    through the lattice. Without loss of generality, we denote the number of 
    empty cells in front of vehicle $i$ as its space gap $g_{s_{i}}$.
  }
  \label{fig:TCABasics}
\end{figure}

In the field of traffic flow modeling, microscopic traffic simulation has always 
been regarded as a time consuming, complex process involving detailed models 
that describe the behavior of individual vehicles. Approximately a decade ago, 
however, new microscopic models were being developed, based on the cellular 
automata programming paradigm from statistical physics. Let us first describe 
the operation of a single-lane traffic cellular automaton as depicted in 
\figref{fig:TCABasics}. We assume $N$ vehicles are driving on a circular lattice 
containing $K$ cells, i.e., periodic boundary conditions (each cell can be 
occupied by at most one vehicle at a time). Time and space are discretized, with 
$\Delta T = 1$~s and $\Delta X = 7.5$~m, leading to a velocity discretization of 
$\Delta V = 27$~km/h. Furthermore, the velocity $v_{i}$ of a vehicle $i$ is 
constrained to an integer in the range $\lbrace 0, \ldots, v_{\mbox{max}} 
\rbrace$, with $v_{\mbox{max}}$ typically 5~cells/s (corresponding to 135~km/h).

Each vehicle $i$ has a space headway $h_{s_{i}}$ and a time headway $h_{t_{i}}$, 
defined as follows:

\begin{eqnarray}
  h_{s_{i}} & = & g_{s_{i}} + l_{i}, \label{eq:SpaceHeadway}\\
  h_{t_{i}} & = & g_{t_{i}} + \rho_{i} \label{eq:TimeHeadway}.
\end{eqnarray}

In these definitions, $g_{s_{i}}$ and $g_{t_{i}}$ denote the space and time gaps 
respectively; $l_{i}$ is the length of a vehicle and $\rho_{i}$ is the occupancy 
time of the vehicle (i.e., the time it `spends' in one cell). Note that in a 
traffic cellular automaton the space headway of a vehicle is always an integer 
number, representing a multiple of the spatial discretization $\Delta X$ in real 
world measurement units. So in a jam, it is taken to be equal to the space the 
vehicle occupies, i.e., $h_{s_{i}} = \one$ cell.

Local interactions between individual vehicles in a traffic stream are modeled 
by means of a rule set. In this paper, we assume that all vehicles have the same 
physical characteristics. The system's state is changed through synchronous 
position updates of all the vehicles, based on a rule set that reflects the 
car-following behavior. Note that most rule sets of TCA models do not use the 
space headway $h_{s_{i}}$ or the space gap $g_{s_{i}}$, but are instead based on 
the number of empty cells $d_{i}$ in front of a vehicle $i$. Keeping equation 
\eqref{eq:SpaceHeadway} in mind, we therefore adopt the convention that, for a 
vehicle $i$ its length $l_{i} = \one$ cell; the resulting TCA models are called 
\emph{single-cell models}. This means that when the vehicle is residing in a 
compact jam, its space headway $h_{s_{i}} = \one$ cell and its space gap is 
consequently $g_{s_{i}} = \zero$ cells. This abstraction gives us a rigorous 
justification to formulate the TCA's update rules more intuitively using space 
gaps.

		\subsection{The stochastic traffic cellular automaton (STCA) model of Nagel and Schreckenberg}

In 1992, Nagel and Schreckenberg proposed a TCA model that was able to reproduce 
several characteristics of real-life traffic flows, e.g., the spontaneous 
emergence of traffic jams \cite{NAGEL:92,NAGEL:95c}. Their model is called the 
\emph{NaSch TCA}, but is more commonly known as the \emph{stochastic traffic 
cellular automaton} (STCA). It explicitly includes a stochastic noise term in 
one of its rules. The STCA then comprises the following three rules (note that 
in Nagel and Schreckenberg's original formulation, they decoupled acceleration 
and braking, resulting in four rules):

\begin{quote}
	\textbf{R1}: \emph{acceleration and braking}\\
		\begin{equation}
		\label{eq:STCAR1}
			v_{i}(t) \leftarrow \mbox{min} \lbrace v_{i}(t - \one) + \one,g_{s_{i}}(t - \one),v_{\mbox{max}} \rbrace,
		\end{equation}

	\textbf{R2}: \emph{randomization}\\
		\begin{equation}
		\label{eq:STCAR2}
			\xi(t) < p \Longrightarrow v_{i}(t) \leftarrow \max \lbrace \zero,v_{i}(t) - \one \rbrace,
		\end{equation}

	\textbf{R3}: \emph{vehicle movement}\\
		\begin{equation}
		\label{eq:STCAR3}
			x_{i}(t) \leftarrow x_{i}(t - \one) + v_{i}(t).
		\end{equation}
\end{quote}

The STCA contains a rule for increasing the speed of a vehicle and braking to 
avoid collisions, i.e., rule R1, equation \eqref{eq:STCAR1}. It furthermore also 
contains rule R2, equation \eqref{eq:STCAR2}, which introduces stochasticity in 
the system. At each discrete time step $t$, a random number $\xi(t) \in 
[\zero,\one[$ is drawn from a uniform distribution. This number is then compared 
with a stochastic noise parameter $p \in [\zero,\one]$ (called the 
\emph{slowdown probability}); as a result, there is a probability of $p$ that a 
vehicle will slow down to $v_{i}(t) - \one$ cells/time step. According to Nagel 
and Schreckenberg, the randomization of rule R2 captures natural speed 
fluctuations due to human behavior or varying external conditions. The rule 
introduces overreactions of drivers when braking, providing the key to the 
formation of spontaneously emerging jams. Finally, rule R3, equation 
\eqref{eq:STCAR3}, allows for the actual movement of vehicles in the system. The 
STCA model is called a \emph{minimal model}, in the sense that all these rules 
are a necessity for mimicking the basic features of real-life traffic flows.

\begin{figure}[!htb]
	\centering
	\includegraphics[width=\figurehalfwidth]{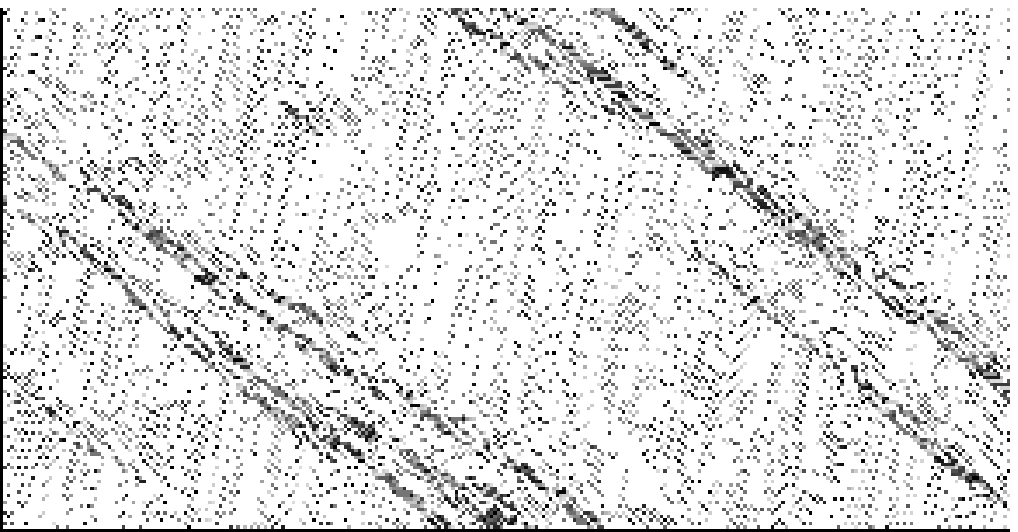}
	\hspace*{0.25cm}
	\includegraphics[width=\figurehalfwidth]{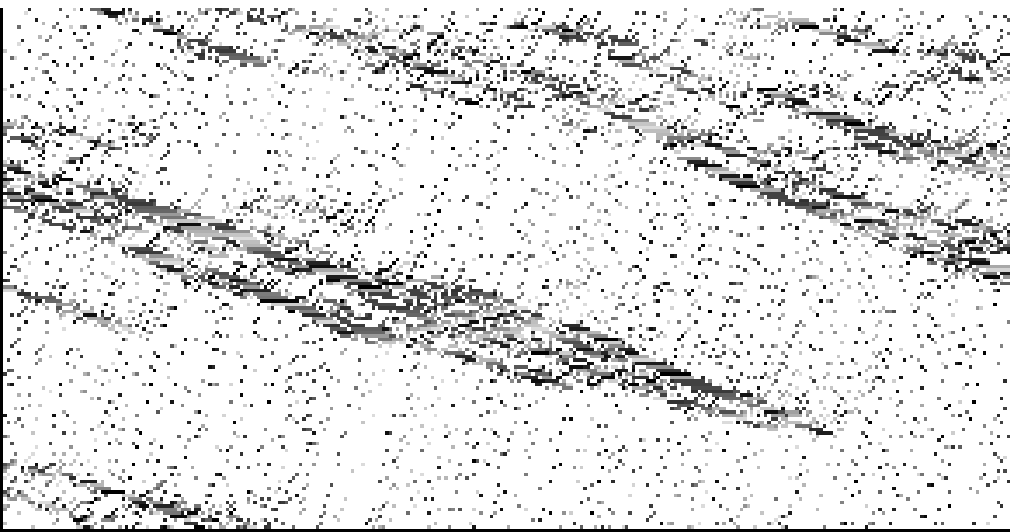}
	\caption{
		Typical time-space diagrams of the STCA model (the time and space axes are 
		oriented left to right, and bottom to top, respectively). The two shown 
		closed-loop lattices each contain 300 cells, with a visible period of 580 
		time steps (each vehicle is represented as a single colored dot). Both 
		diagrams have a global density of $k =$~0.2 vehicles/cell. \emph{Left:} the 
		evolution of the system for $p =$~0.1. \emph{Right:} the evolution of the 
		system, but now for $p = $~0.5. The effects of the randomization rule R2 are 
		clearly visible in both diagrams, as there occur many unstable artificial 
		phantom mini-jams. Furthermore, the speed $w$ of the backward propagating 
		kinematic waves decreases with an increasing $p$.
	}
	\label{fig:STCATimeSpaceDiagrams}
\end{figure}

To get an intuitive feeling for the STCA's system dynamics, we have provided two 
time-space diagrams in \figref{fig:STCATimeSpaceDiagrams}. Both diagrams show 
the evolution for a global density of $k =$~0.2 vehicles/cell, but with $p$ set 
to 0.1 for the left diagram, and $p = $~0.5 for the right diagram. As can be 
seen in both diagrams, the randomization in the model gives rise to many 
unstable artificial phantom mini-jams. The downstream fronts of these jams smear 
out, forming \emph{unstable interfaces} \cite{NAGEL:03}. This is a direct result 
of the fact that the intrinsic noise (as embodied by $p$) in the STCA model is 
too strong: a jam can always form at \emph{any} density, meaning that breakdown 
can (and will) occur, even in the free-flow traffic regime. For low enough 
densities however, these jams can vanish as they are absorbed by vehicles with 
sufficient space headways, or by new jams in the system \cite{KRAUSS:99}. It has 
been experimentally shown that below the critical density, these jams have 
finite life times with a cut-off that is about $\five \times \ten^{\five}$ time 
steps and independent of the lattice size. When the critical density is crossed, 
these long-lived jams evolve into jams with an infinite life time, i.e., they 
will survive for an infinitely long time 
\cite{NAGEL:94b,NAGEL:95c,SCHADSCHNEIDER:99b}.

		\subsection{The macroscopic first-order traffic flow model of Lighthill, Whitham, and Richards (LWR)}
		\label{sec:LWRModelDescription}

Considering traffic as an inviscid but compressible fluid, it follows from this 
assumption, that densities $k$, space-mean speeds $\overline v_{s}$, and flows 
$q$ are defined as continuous variables, in each point in time and space, hence 
leading to the names of \emph{continuum models}, \emph{fluid-dynamic models}, or 
\emph{macroscopic models}.

The first aspect of such a fluid-dynamic description of traffic flow, consists 
of a \emph{scalar conservation law}; `scalar' because it is a first-order 
partial differential equation (PDE). A typical derivation can be found in 
\cite{GARTNER:97} and \cite{JUNGEL:02}: the derivation is based on considering a 
road segment with a finite length on which no vehicles appear or disappear other 
than the ones that enter and exit it. After taking the infinitesimal limit, this 
will result in a PDE that expresses the interplay between continuous densities 
and flows on a local scale:

\begin{equation}
\label{eq:FirstOrderConservationLaw}
	\frac{\partial k(t,x)}{\partial t} + \frac{\partial q(t,x)}{\partial x} = \zero,
\end{equation}

with the density $k$ and flow $q$ dynamically (i.e., time varying) defined over 
a single spatial dimension. Lighthill and Whitham were among the first to 
develop such a traffic flow model in 1955 \cite{LIGHTHILL:55}. One year later, 
Richards independently derived the same fluid-dynamic model \cite{RICHARDS:56}, 
albeit in a slightly different form. Because of the nearly simultaneous and 
independent development of the theory, the model has become known as the 
\emph{LWR model}, after the initials of its inventors who receive the credit. In 
some texts, the model is also referred to as the \emph{hydrodynamic model}, or 
the \emph{kinematic wave model} (KWM), attributed to the fact that the model's 
solution is based on characteristics, which are called kinematic waves (e.g., 
shock waves).

Crucial to their approach, was a fundamental hypothesis, essentially stating 
that flow is a function of density, i.e., there exists a $q_{e}(k(t,x))$ 
equilibrium relationship, more commonly known as the \emph{fundamental diagram} 
\cite{HAIGHT:63}. Central to their theory, Lighthill and Whitham assumed that 
the fundamental hypothesis holds at all traffic densities, not just for 
light-density traffic but also for congested traffic conditions. 

In order to solve equation \eqref{eq:FirstOrderConservationLaw}, we also need 
the \emph{fundamental relation of traffic flow theory}, which relates the 
macroscopic traffic flow characteristics density $k$ (vehicles/kilometre), 
space-mean speed $\overline v_{s}$ (kilometres/hour), and flow $q$ 
(vehicles/hour) to each other as follows:

\begin{equation}
\label{eq:FundamentalRelation}
	q = k~\overline v_{s}.
\end{equation}

In general however, there are two restrictions, i.e., the relation is only valid 
for (1) continuous variables, or smooth approximations of them, and (2) traffic 
composed of substreams (e.g., slow and fast vehicles) which comply to the 
following two assumptions:

\begin{quote}
	\begin{description}
		\item[\textbf{Homogeneous traffic}]~

			There is a homogeneous composition of the traffic substream (i.e., the 
			same type of vehicles).\\
		\item[\textbf{Stationary traffic}]~

			When observing the traffic substream at different times and locations, it 
			`looks the same'. Putting it a bit more quantitatively, all the vehicles' 
			trajectories in a time-space diagram should be parallel and equidistant 
			\cite{DAGANZO:97}.
	\end{description}
\end{quote}

The latter of the above two conditions, is also referred to as traffic operating 
in a \emph{steady state} or at \emph{equilibrium}. Employing the fundamental 
diagram $q_{e}(k)$, relates the two dependent variables in equation 
\eqref{eq:FirstOrderConservationLaw} to each other, thereby making it possible 
to solve the partial differential equation. Thus, reconsidering equation 
\eqref{eq:FirstOrderConservationLaw}, taking into account the fundamental 
diagram, the conservation law is now expressed as:

\begin{equation}
\label{eq:FirstOrderPDEConservationLaw}
	k_{t} + q_{e}(k)_{x} = \zero,
\end{equation}

in which we introduced the standard differential calculus notation for PDEs. 
Recognizing the fundamental relation of traffic flow theory 
\eqref{eq:FundamentalRelation}, the conservation law 
\eqref{eq:FirstOrderPDEConservationLaw} can also be cast in a non-linear wave 
equation, using the chain rule for differentiation \cite{GARTNER:97}:

\begin{equation}
\label{eq:FirstOrderPDEIVP}
	k_{t} + \frac{dq_{e}(k)}{dk} k_{x} = \zero.
\end{equation}

Analytically solving the previous equation using the method of characteristics, 
results in shock waves that travel with speeds equal to:

\begin{equation}
	w = \frac{dq_{e}(k)}{dk},
\end{equation}

i.e., the tangent to the $q_{e}(k)$ fundamental diagram. As a consequence, 
solutions, being the characteristics, of equation \eqref{eq:FirstOrderPDEIVP} 
have the following form:

\begin{equation}
	k(t,x) = k(x - wt),
\end{equation}

with the observation that the density is constant along such a characteristic. 
Whenever in the solution of the conservation equation, two of its 
characteristics intersect, the density takes on two different values (each one 
belonging to a single characteristic). As this mathematical quirk is physically 
impossible, the so-called entropy solution states that both characteristics 
terminate and breed a \emph{shock wave}; as such, these shock waves form 
boundaries that discontinuously separate densities, flows, and space-mean speeds 
\cite{GARTNER:97}. The speed of such a shock wave is related to the following 
ratio \cite{PIPES:64}:

\begin{equation}
\label{eq:FirstOrderShockWaveSpeed}
	w_{\mbox{shock}} = \frac{\Delta q}{\Delta k},
\end{equation}

with $\Delta q = q_{u} - q_{d}$ and $\Delta k = k_{u} - k_{d}$ the relative 
difference in flows, respectively densities, up- and downstream of the shock 
wave.

Note that going from a low to a high density regime typically results in a shock 
wave, whereas the reverse transition is accompanied by an emanation of a 
\emph{fan of characteristics} (also called \emph{expansion}, 
\emph{acceleration}, or \emph{rarefaction waves}). In shock wave theory, the 
densities on either side of a shock are well defined (i.e., unique solutions 
exist); along the shock wave however, the density jumps discontinuously from one 
value to another.

Because of the well-defined properties of the LWR model, it is possible to 
derive analytical solutions to certain types of problems. These solutions can 
even be drawn graphically in a time-space diagram, thereby clearly sho\-wing the 
evolution of first-order macroscopic traffic flow characteristics (e.g., the 
speed of backward-travelling jams, \ldots). Besides the previous analytic 
derivation of a solution to the conservation law expressed as a PDE, it is also 
possible to treat the problem numerically. Converting the PDEs into finite 
difference equations (FDEs), and solving them numerically stable, can be done by 
casting the LWR model in the context of \emph{Godunov FDE methods}, allowing for 
arbitrary $q_{e}(k)$ fundamental diagrams \cite{DAGANZO:95b,LEBACQUE:96}. As 
such, the LWR model is very appealing because it provides an attractive means 
for gaining insight into the primary characteristics of the evolution of traffic 
flows. As a surplus, the model has the benefits of the ability to efficiently 
construct analytical solutions, as well as the existence of thoroughly-developed 
and understood numerical schemes.

	\section{Implicitly incorporating the STCA's stochasticity}
	\label{sec:InitialMethodology}

As mentioned in the introduction, we reconsider the STCA and LWR models, taking 
a different approach at studying their relation. Our main goal is to provide a 
means for implicitly incorporating the STCA's stochasticity into the LWR model. 
To this end, we provide a practical methodology for specifying the fundamental 
diagram to the LWR model. Assuming that a \emph{stationarity condition} holds on 
the STCA's rules, we incorporate the STCA's stochasticity directly into the 
LWR's fundamental diagram.

Relating both the STCA and the LWR models is now done using a simple two-step 
approach, in which we first rewrite the STCA's rules into a single rule, leading 
to a set of \emph{linear inequalities}. These constraints can be considered as a 
$\overline v_{s_{e}}(\overline h_{s})$ fundamental diagram (see e.g., the left 
part in \figref{fig:DerivingStationaryFundamentalDiagrams}). This latter diagram 
can then be converted into an equivalent triangular flow versus density 
$q_{e}(k)$ fundamental diagram.

		\subsection{Rewriting the STCA's rule set}

Considering a vehicle's average speed, the STCA's rules R1 and R2, equations 
\eqref{eq:STCAR1} and \eqref{eq:STCAR2} respectively, state that a vehicle slows 
down with probability $p$, and that it does not slow down with probability $\one 
- p$. As such, they can be rewritten into the following single rule that is 
expressed in \emph{continuous} speeds and space gaps:

\begin{eqnarray}
	v_{i}(t) & \leftarrow & p \cdot \mbox{min} \lbrace v_{i}(t - \one)~\cancel{+ \one}~~\cancel{- \one}, g_{s_{i}}(t - \one) - \one,\nonumber\\
	         &            & v_{\mbox{max}} - \one \rbrace + (\one - p) \cdot \mbox{min} \lbrace v_{i}(t - \one) + \one,\nonumber\\
	         &            & g_{s_{i}}(t - \one), v_{\mbox{max}} \rbrace,\label{eq:STCASingleRule}
\end{eqnarray}

with $v_{i}(t) \leftarrow \mbox{max} \lbrace 0, v_{i}(t) \rbrace$. Furthermore, 
the following two algebraic relations always hold:

\begin{eqnarray}
	a \cdot \mbox{min} \lbrace b, c \rbrace
		& = &
		\mbox{min} \lbrace ab, ac \rbrace,\label{eq:MinAlgebraRule1}\\
	\mbox{min} \lbrace a, b \rbrace + \mbox{min} \lbrace c, d \rbrace
		& = &
		\mbox{min} \lbrace a + c, a + d,\nonumber\\
		&   &
		\qquad b + c, b + d \rbrace.\label{eq:MinAlgebraRule2}
\end{eqnarray}

Applying relation \eqref{eq:MinAlgebraRule1} to our rule 
\eqref{eq:STCASingleRule}, yields the following result:

\begin{eqnarray}
	v_{i}(t)
		& \leftarrow &
		\mbox{min} \lbrace p v_{i}(t - \one), p (g_{s_{i}}(t - \one) - \one), p (v_{\mbox{max}} - \one) \rbrace \nonumber\\
	  &            &
	  + \mbox{min} \lbrace (\one - p) (v_{i}(t - \one) + \one), (\one - p) g_{s_{i}}(t - \one),\nonumber\\
	  &            &
	  \qquad~~ (\one - p) v_{\mbox{max}} \rbrace.
\end{eqnarray}

Using relation \eqref{eq:MinAlgebraRule2} to this result, allows us to obtain a 
formulation with a single minimum-operator:

\begin{eqnarray}
	v_{i}(t) \leftarrow \mbox{min} \lbrace
	& & p v_{i}(t - \one) + (\one - p) (v_{i}(t - \one) + \one),\nonumber\\
	& & p v_{i}(t - \one) + (\one - p) g_{s_{i}}(t - \one),\nonumber\\
	& & p v_{i}(t - \one) + (\one - p) v_{\mbox{max}},\nonumber\\
	& & p (g_{s_{i}}(t - \one) - \one) + (\one - p) (v_{i}(t - \one) + \one),\nonumber\\
	& & p (g_{s_{i}}(t - \one) - \one) + (\one - p) g_{s_{i}}(t - \one),\nonumber\\
	& & p (g_{s_{i}}(t - \one) - \one) + (\one - p) v_{\mbox{max}},\nonumber\\
	& & p (v_{\mbox{max}} - \one) + (\one - p) (v_{i}(t - \one) + \one),\nonumber\\
	& & p (v_{\mbox{max}} - \one) + (\one - p) g_{s_{i}}(t - \one),\nonumber\\
	& & p (v_{\mbox{max}} - \one) + (\one - p) v_{\mbox{max}} \rbrace.
\end{eqnarray}

Expanding all the terms between parentheses gives the following result:

\begin{eqnarray}
	v_{i}(t) \leftarrow \mbox{min} \lbrace
	& & \cancel{p v_{i}(t - \one)} + v_{i}(t - \one) + \one - \cancel{p v_{i}(t - \one)} - p,\nonumber\\
	& & p v_{i}(t - \one) + g_{s_{i}}(t - \one) - p g_{s_{i}}(t - \one),\nonumber\\
	& & p v_{i}(t - \one) + v_{\mbox{max}} - p v_{\mbox{max}},\nonumber\\
	& & p g_{s_{i}}(t - \one) - p + v_{i}(t - \one) + \one - p v_{i}(t - \one) - p,\nonumber\\
	& & \cancel{p g_{s_{i}}(t - \one)} - p + g_{s_{i}}(t - \one) - \cancel{p g_{s_{i}}(t - \one)},\nonumber\\
	& & p g_{s_{i}}(t - \one) - p + v_{\mbox{max}} - p v_{\mbox{max}},\nonumber\\
	& & p v_{\mbox{max}} - p + v_{i}(t - \one) + \one - p v_{i}(t - \one) - p,\nonumber\\
	& & p v_{\mbox{max}} - p + g_{s_{i}}(t - \one) - p g_{s_{i}}(t - \one),\nonumber\\
	& & \cancel{p v_{\mbox{max}}} - p + v_{\mbox{max}} - \cancel{p v_{\mbox{max}}} \rbrace.
\end{eqnarray}

And finally, regrouping for $p$ yields:

\begin{eqnarray}
	v_{i}(t) \leftarrow \mbox{min} \lbrace
	& & v_{i}(t - \one) + \one - p,\nonumber\\
	& & p (v_{i}(t - \one) - g_{s_{i}}(t - \one)) + g_{s_{i}}(t - \one),\nonumber\\
	& & p (v_{i}(t - \one) - v_{\mbox{max}}) + v_{\mbox{max}},\nonumber\\
	& & p (g_{s_{i}}(t - \one) - v_{i}(t - \one) - \two) + v_{i}(t - \one) + \one,\nonumber\\
	& & g_{s_{i}}(t - \one) - p,\nonumber\\
	& & p (g_{s_{i}}(t - \one) - v_{\mbox{max}} - \one) + v_{\mbox{max}},\nonumber\\
	& & p (v_{\mbox{max}} - v_{i}(t - \one) - \two) + v_{i}(t - \one) + \one,\nonumber\\
	& & p (v_{\mbox{max}} - g_{s_{i}}(t - \one) - \one) + g_{s_{i}}(t - \one),\nonumber\\
	& & v_{\mbox{max}} - p \rbrace.\label{eq:STCASingleRuleRegrouped}
\end{eqnarray}

If we now assume traffic is \emph{stationary} (see e.g., Daganzo's description 
of stationary traffic in section \ref{sec:LWRModelDescription}), then we can 
assert that the state of a vehicle at time $t$ is the same as its state at time 
$t - \one$, i.e., $v_{i}(t) = v_{i}(t - \one)$ and $g_{s_{i}}(t) = g_{s_{i}}(t - 
\one)$. As a result, equation \eqref{eq:STCASingleRuleRegrouped} gets 
transformed into the following set of \emph{linear inequalities} that express 
constraints on the relations between $v_{i}(t)$, $g_{s_{i}}(t)$, $p$, and 
$v_{\mbox{max}}$:

\begin{eqnarray}
	\cancel{v_{i}(t)} + \one - p \geq \cancel{v_{i}(t)}
		& & \quad \mbox{(C1)},\nonumber\\
	p (v_{i}(t) - g_{s_{i}}(t)) + g_{s_{i}}(t) \geq v_{i}(t)
		& & \quad \mbox{(C2)},\nonumber\\
	p (v_{i}(t) - v_{\mbox{max}}) + v_{\mbox{max}} \geq v_{i}(t)
		& & \quad \mbox{(C3)},\nonumber\\
	p (g_{s_{i}}(t) - v_{i}(t) - \two) + \cancel{v_{i}(t)} + \one \geq \cancel{v_{i}(t)}
		& & \quad \mbox{(C4)},\nonumber\\
	g_{s_{i}}(t) - p \geq v_{i}(t)
		& & \quad \mbox{(C5)},\nonumber\\
	p (g_{s_{i}}(t) - v_{\mbox{max}} - \one) + v_{\mbox{max}} \geq v_{i}(t)
		& & \quad \mbox{(C6)},\nonumber\\
	p (v_{\mbox{max}} - v_{i}(t) - \two) + \cancel{v_{i}(t)} + \one \geq \cancel{v_{i}(t)}
		& & \quad \mbox{(C7)},\nonumber\\
	p (v_{\mbox{max}} - g_{s_{i}}(t) - \one) + g_{s_{i}}(t) \geq v_{i}(t)
		& & \quad \mbox{(C8)},\nonumber\\
	v_{\mbox{max}} - p \geq v_{i}(t)
		& & \quad \mbox{(C9)}.\nonumber
\end{eqnarray}

Let us now examine each of these nine constraints C1 through C9.

\begin{itemize}
	\item[$\bullet$] Constraint C1 states that $\one - p \geq \zero$, i.e., $p \leq \one$. 
	This logically follows from the STCA's condition that $p \in [\zero,\one]$.

	\item[$\bullet$] Constraint C2 states that $p (v_{i}(t) - g_{s_{i}}(t)) + g_{s_{i}}(t) 
	\geq v_{i}(t)$, i.e., $g_{s_{i}}(t) \cancel{(\one - p)} \geq v_{i}(t) 
	\cancel{(\one - p)}$. This corresponds to $v_{i}(t) \leq g_{s_{i}}(t)$, which 
	states that vehicles strive for collision-free driving.

	\item[$\bullet$] Constraint C3 states that $p (v_{i}(t) - v_{\mbox{max}}) + 
	v_{\mbox{max}} \geq v_{i}(t)$, i.e., $v_{\mbox{max}} \cancel{(\one - p)} \geq 
	v_{i}(t) \cancel{(\one - p)}$. This corresponds to $v_{i}(t) \leq 
	v_{\mbox{max}}$, which logically follows from the STCA's condition that 
	$v_{i}(t) \in \lbrace \zero, \ldots, v_{\mbox{max}} \rbrace$.

	\item[$\bullet$] Constraint C4 states that $p (g_{s_{i}}(t) - v_{i}(t) - \two) + \one 
	\geq \zero$, i.e., $v_{i}(t) \leq g_{s_{i}}(t) - \two + \frac{\one}{p}$ (for 
	$p \neq \zero$).

	\item[$\bullet$] Constraint C5 states that $g_{s_{i}}(t) - p \geq v_{i}(t)$, i.e., 
	$v_{i}(t) \leq g_{s_{i}}(t) - p$, which is a more stringent constraint than C2 
	and C4.

	\item[$\bullet$] Constraint C6 states that $p (g_{s_{i}}(t) - v_{\mbox{max}} - \one) + 
	v_{\mbox{max}} \geq v_{i}(t)$, i.e., $v_{i}(t) \leq v_{\mbox{max}} (\one - p) 
	+ p (g_{s_{i}}(t) - \one)$. 

	\item[$\bullet$] Constraint C7 states that $p (v_{\mbox{max}} - v_{i}(t) - \two) + \one 
	\geq \zero$, i.e., $v_{i}(t) \leq v_{\mbox{max}} - \two + \frac{\one}{p}$ (for 
	$p \neq \zero$). 

	\item[$\bullet$] Constraint C8 states that $p (v_{\mbox{max}} - g_{s_{i}}(t) - \one) + 
	g_{s_{i}}(t) \geq v_{i}(t)$, i.e., $v_{i}(t) \leq g_{s_{i}}(t) (\one - p) + p 
	(v_{\mbox{max}} - \one)$. 

	\item[$\bullet$] Constraint C9 states that $v_{\mbox{max}} - p \geq v_{i}(t)$, i.e., 
	$v_{i}(t) \leq v_{\mbox{max}} - p$, which is a more stringent constraint than 
	C3 and C7.
\end{itemize}

Taking the previous considerations into account, we can see that constraints C1, 
C2, and C3 are always satisfied. The remaining three pairs of similar 
constraints on the relations between $v_{i}(t)$, $g_{s_{i}}(t)$, $p$, and 
$v_{\mbox{max}}$, are the following: constraints C5 and C9, C4 and C7, and C6 
and C8.

\begin{figure}[!htb]
	\centering
	\psfrag{vi(t)}[][]{\psfragstyle{$v_{i}(t)$}}
	\psfrag{gsi(t)}[][]{\psfragstyle{$g_{s_{i}}(t)$}}
	\psfrag{1}[][]{\psfragstyle{1}}
	\psfrag{vmax}[][]{\psfragstyle{$v_{\mbox{max}}$}}
	\psfrag{vmax-1}[][l]{\psfragstyle{$v_{\mbox{max}} - \one$}}
	\psfrag{vmax+1}[][l]{\psfragstyle{$v_{\mbox{max}} + \one$}}
	\psfrag{C6}[][]{\psfragstyle{C6}}
	\psfrag{C8}[][]{\psfragstyle{C8}}
	\includegraphics[width=\figurewidth]{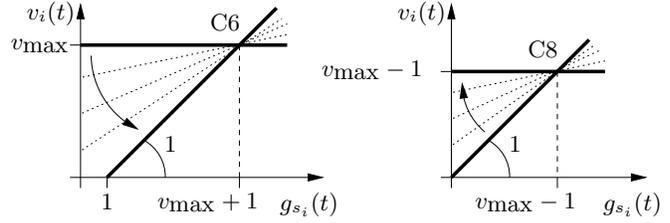}
	\caption{
		A visual representation of the constraints C6 and C8. \emph{Left:} as $p = 
		\zero \rightarrow \one$, C6 changes from a horizontal line at $v_{i}(t) = 
		v_{\mbox{max}}$, to a slanted line with a slope of $+\one$, intercepting the 
		horizontal and vertical axes at $+\one$ and $-\one$, respectively. 
		\emph{Right:} at the same time, constraint C8 changes from a slanted line 
		with a slope of $+\one$, passing through the origin, into a horizontal line 
		at $v_{i}(t) = v_{\mbox{max}} - \one$.
	}
	\label{fig:Constraints6And8}
\end{figure}

In order to gain insight into the more difficult constraints C6 and C8, we first 
rewrite them as follows:

\begin{eqnarray}
	\mbox{(C6)} & & v_{i}(t) \leq \underbrace{p}_{\mbox{slope}} g_{s_{i}}(t) + \underbrace{(\one - p) v_{\mbox{max}} - p}_{\mbox{intercept}},\nonumber\\
	\mbox{(C8)} & & v_{i}(t) \leq \underbrace{(\one - p)}_{\mbox{slope}} g_{s_{i}}(t) + \underbrace{p (v_{\mbox{max}} - \one)}_{\mbox{intercept}},\nonumber
\end{eqnarray}

where we have separated the terms containing $g_{s_{i}}(t)$. Plotting the speed 
$v_{i}(t)$ versus the space gap $g_{s_{i}}(t)$ in \figref{fig:Constraints6And8}, 
allows us to more easily interpret the combined effects of these two 
constraints. On the one hand, if we continuously change $p = \zero \rightarrow 
\one$, then constraint C6 goes from a horizontal line at $v_{i}(t) = 
v_{\mbox{max}}$, to a slanted line with a slope of $+\one$, intercepting the 
horizontal and vertical axes at $+\one$ and $-\one$, respectively. In all cases, 
the point at ($v_{\mbox{max}} + \one$,$v_{\mbox{max}}$) remains invariant. On 
the other hand, changing $p = \zero \rightarrow \one$ turns constraint C8 from a 
slanted line with a slope of $+\one$, passing through the origin, into a 
horizontal line at $v_{i}(t) = v_{\mbox{max}} - \one$. In all cases, the point 
at ($v_{\mbox{max}} - \one$,$v_{\mbox{max}} - \one$) remains invariant.

		\subsection{Deriving the fundamental diagram}

The next step of our approach, considers the most determining linear 
inequalities C5, C6, C8, and C9 as boundaries in a $\overline 
v_{s_{e}}(\overline g_{s})$ fundamental diagram. As such, we note the following 
observations:

\begin{itemize}
	\item[$\bullet$] \textbf{Increasing the slowdown probability $p$, holding $v_{\mbox{max}}$ constant}:
	\begin{itemize}
		\item The average speed $\overline v_{\mbox{ff}}$ in the free-flow regime 
		decreases towards $v_{\mbox{max}} - p$.
		\item The transition point at the critical space gap $g_{s_{c}}$ remains
		invariant.
		\item The space gap $g_{s_{j}}$, corresponding to the jam density, 
		increases.
	\end{itemize}
	\item[$\bullet$] \textbf{Decreasing the maximum speed $v_{\mbox{max}}$, holding $p$ constant}:
	\begin{itemize}
		\item The average speed $\overline v_{\mbox{ff}}$ in the free-flow regime 
		decreases towards $v_{\mbox{max}} - p$.
		\item The transition point at the critical space gap $g_{s_{c}}$ decreases.
		\item The space gap $g_{s_{j}}$, corresponding to the jam density, remains 
		invariant.
	\end{itemize}
\end{itemize}

From equation \eqref{eq:SpaceHeadway} from traffic flow theory, it follows that 
a vehicle's space headway $h_{s}$ is equal to its space gap $g_{s}$ (i.e., the 
distance between the vehicle's frontal bumper and the one of its direct frontal 
leader), plus the vehicle's own length $l$. As such, the derived $\overline 
v_{s_{e}}(\overline g_{s})$ fundamental diagram can be converted into a 
$\overline v_{s_{e}}(\overline h_{s})$ fundamental diagram. Because we 
originally started from a single-cell TCA model (i.e., the STCA model with all 
vehicles having the same unit length), we can use our convention which states 
that a vehicle's length $l_{i} \geq \one~\mbox{cell} \propto \Delta X$ (see our 
discussion at the end of section \ref{sec:TCAModels} for more details).

\begin{figure}[!htb]
	\centering
	\psfrag{hs}[][]{\tiny{$\overline h_{s}$}}
	\psfrag{vs}[][]{\tiny{$\overline v_{s}$}}
	\psfrag{k}[][]{\tiny{$k$}}
	\psfrag{DeltaX}[][]{\tiny{$\Delta X$}}
	\psfrag{DeltaX-1}[][]{\tiny{$\Delta X^{-\one}$}}
	\psfrag{kc-1}[][]{\tiny{$k_{c}^{-\one}$}}
	\psfrag{kj-1}[][]{\tiny{$k_{j}^{-\one}$}}
	\psfrag{kc}[][]{\tiny{$k_{c}$}}
	\psfrag{kj}[][]{\tiny{$k_{j}$}}
	\psfrag{q}[][]{\tiny{$q$}}
	\psfrag{qcap}[][]{\tiny{$q_{\mbox{cap}}$}}
	\psfrag{vmax}[][]{\tiny{$v_{\mbox{max}}$}}
	\psfrag{vmax-p}[][]{\tiny{$v_{\mbox{max}} - p$}}
	\psfrag{deterministic}[][]{\tiny{deterministic}}
	\psfrag{stochastic}[][]{\tiny{stochastic}}
	\includegraphics[width=\figurewidth]{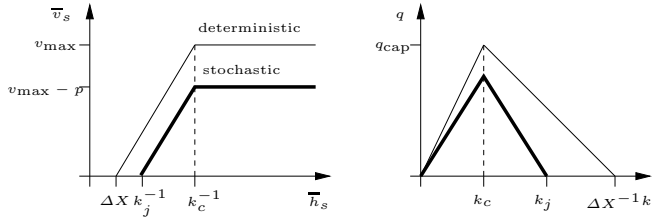}
	\caption{
		\emph{Left:} deriving a stationary $\overline v_{s_{e}}(\overline h_{s})$ 
		fundamental diagram from the STCA's constraints C1 -- C9. The stochastic 
		diagram has a higher inverse jam density, but the same inverse critical 
		density as its deterministic counterpart (for the same $v_{\mbox{max}}$). 
		\emph{Right:} an equivalent triangular $q_{e}(k)$ fundamental diagram.
	}
	\label{fig:DerivingStationaryFundamentalDiagrams}
\end{figure}

Let us now consider the relation between the macroscopic traffic flow 
characteristic density $k$ and the microscopic characteristic average space 
headway $\overline h_{s}$, i.e., $\overline h_{s} = k^{-\one}$ 
\cite{WARDROP:52,DAGANZO:97}. This allows us to effectively transform the 
$\overline v_{s_{e}}(\overline h_{s})$ fundamental diagram into a $\overline 
v_{s_{e}}(k)$ fundamental diagram. As can be seen in the left part of 
\figref{fig:DerivingStationaryFundamentalDiagrams}, increasing the stochasticity 
leads to the same observations that we previously mentioned. Finally, using the 
fundamental relation of traffic flow theory \eqref{eq:FundamentalRelation} (see 
section \ref{sec:LWRModelDescription}), our constraints are transformed into an 
equivalent \emph{triangular} $q_{e}(k)$ fundamental diagram. Applying this 
technique results in the following analytical expressions for the parameters of 
the LWR's fundamental diagram:

\begin{eqnarray}
	\overline v_{\mbox{ff}} & = & (v_{\mbox{max}} - p) ~ \frac{\Delta X}{\Delta T} ~ \mbox{3.6},\label{eq:LWRFDExpressionVFF}\\
	k_{\mbox{crit}}         & = & \frac{\mbox{1000}}{(v_{\mbox{max}} + l) ~ \Delta X},\\
	k_{\mbox{jam}}          & = & \frac{\mbox{1000}}{(l + p) ~ \Delta X},
\end{eqnarray}

with $l = \one$ cell as the length of all vehicles in the single-cell STCA 
model. The capacity flow is calculated using the fundamental relation 
\eqref{eq:FundamentalRelation}, resulting in the following expression:

\begin{equation}
\label{eq:LWRFDExpressionQCAP}
	q_{\mbox{cap}} = k_{\mbox{crit}}~\overline v_{\mbox{ff}}.
\end{equation}

As is visible in the right part of 
\figref{fig:DerivingStationaryFundamentalDiagrams}, an increase of the 
stochasticity leads to a lower capacity flow $q_{\mbox{cap}}$, an invariant 
critical density $k_{c}$, and a smaller jam density $k_{j}$.

In conclusion, we note how rewriting the STCA's rule set allowed us to obtain a 
stationary triangular $q_{e}(k)$ fundamental diagram. This fundamental diagram, 
which implicitly incorporates the STCA's stochasticity, can then be specified as 
a parameter to the macroscopic first-order LWR model of section 
\ref{sec:LWRModelDescription}. Finally note that the shape of the derived 
fundamental diagram is dictated by the inequality constraints C1 -- C9. As such, 
it actually represents an `outer envelope', that is to say, all possible 
fundamental diagrams lie beneath this envelope. This includes curved fundamental 
diagrams, more generally piecewise linear fundamental diagrams, et cetera.

	\section{Application to an illustrative case study}
	\label{sec:ApplicationToACaseStudy}

After deriving a relation between the STCA and LWR models by means of the 
process explained in the previous section \ref{sec:InitialMethodology}, we now 
apply our methodology to a small case study. We first describe the setup of the 
test scenario, after which we interpret and discuss our obtained results.

		\subsection{Description of the case study}

\begin{figure}[!htb]
	\centering
	\psfrag{0 km}[][]{\scriptsize{\emph{0~km}}}
	\psfrag{11 km}[][]{\scriptsize{\emph{11~km}}}
	\psfrag{17 km}[][]{\scriptsize{\emph{17~km}}}
	\psfrag{22 km}[][]{\scriptsize{\emph{22~km}}}
	\psfrag{27 km/h}[][]{\tiny{\textbf{27~km/h}}}
	\psfrag{135 km/h}[][]{\tiny{\textbf{135~km/h}}}
	\psfrag{traffic}[][]{\scriptsize{traffic}}
	\psfrag{demand}[][]{\scriptsize{demand}}
	\psfrag{Section A}[][]{\scriptsize{Section $A$}}
	\psfrag{Section B}[][]{\scriptsize{Section $B$}}
	\psfrag{Section C}[][]{\scriptsize{Section $C$}}
	\includegraphics[width=\figurewidth]{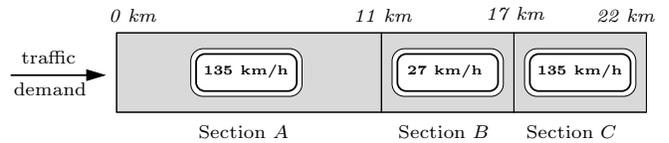}
	\caption{
		The single-lane road of the case study we consider, consisting of three 
		consecutive segments $A$, $B$, and $C$. Assuming temporal and spatial 
		discretizations of $\Delta T = \one$~s and $\Delta X =$~7.5~m, respectively, 
		segment $A$ is composed of 1500 cells, while segments $B$ and $C$ are each 
		composed of 750 cells. The maximum speed for segments $A$ and $C$ is 
		$v_{\mbox{max}} = \five$~cells/time step, whereas it is $v_{\mbox{max}} = 
		\one$~cell/time step for segment $B$.
	}
	\label{fig:TestScenarioDescription}
\end{figure}

The case study we consider, is modeled as a single-lane road that has a middle 
section with a reduced maximum speed (corresponding to e.g., an elevation, a 
speed limit, \ldots). This road consists of three consecutive segments $A$, $B$, 
and $C$, as depicted in \figref{fig:TestScenarioDescription}, whereby vehicles 
enter the road at segment $A$, travel through segment $B$, and exit it at the 
end of segment $C$. For the STCA, we assume a temporal and spatial 
discretization of $\Delta T = \one$~s and $\Delta X =$~7.5~m, respectively. The 
first road segment $A$ then consists of 1500 cells (11.25~km), while the second 
and third segments $B$ and $C$ each consist of 750 cells (i.e., each 
approximately 5.6~km long). The maximum speed for segments $A$ and $C$ is 
$v_{\mbox{max}}^{A,C} = \five$~cells/time step, whereas it is 
$v_{\mbox{max}}^{B} = \one$~cell/time step for segment $B$. The capacity flows 
for all three segments are denoted as $q_{\mbox{cap}}^{A,C}$ and 
$q_{\mbox{cap}}^{B}$.

\begin{figure}[!htb]
	\centering
	\framebox{\includegraphics[width=\figurewidth]{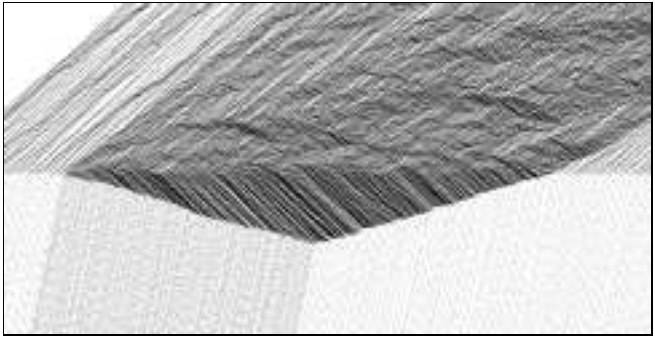}}
	\caption{
		A close up of the individual vehicle trajectories for the STCA in a 
		time-space diagram, near the border between segments $A$ 
		($v_{\mbox{max}}^{A,C} = \five$ cells/time step) and $B$ 
		($v_{\mbox{max}}^{B} = \one$ cell/time step), for $p =$~0.1 everywhere in 
		the system. We can see the formation and dissolution of an upstream growing 
		congested region at the end of segment $A$, related to the short traffic 
		burst.
	}
	\label{fig:TCATrajectoriesCloseup}
\end{figure}

This road is simulated using both the STCA and the LWR model, each time for 3000 
time steps. As for the boundary conditions, we assume an overall inflow of 
$q_{\mbox{cap}}^{B} / \two$, except from time step 200 until time step 600, 
where we have created a short \emph{traffic burst} of increased demand, with an 
inflow of $(q_{\mbox{cap}}^{A,C} + q_{\mbox{cap}}^{B}) / \two$. 
\figref{fig:TCATrajectoriesCloseup} shows a close up of the individual vehicle 
trajectories for the STCA in a time-space diagram, near the border between 
segments $A$ and $B$. As can be seen from the trajectories, heavy congestion 
sets in and flows upstream into segment $A$, where it starts to dissolve at the 
end of the traffic burst. The result is a typical triangular-shaped region that 
contains a queue of slow-moving vehicles (the backward propagating waves are 
clearly distinguished as the pattern of parallel black and white stripes).

\setlength{\fboxsep}{0pt}
\setlength{\fboxsep}{\tempfboxsep}

Applying our previously discussed methodology, we construct a stationary 
triangular $q_{e}(k)$ fundamental diagram. Its parameters are calculated by 
means of equations \eqref{eq:LWRFDExpressionVFF} -- 
\eqref{eq:LWRFDExpressionQCAP}. The results are listed in 
\tableref{table:LWRCalculatedFDParameters}, with the TCA's parameters expressed 
in cells/time step, vehicles/cell, and vehicles/time step, respectively, and the 
LWR's parameters expressed in kilometres/hour, vehicles/\-kilometre, and 
vehicles/hour, respectively.

\begin{table}[!htb]
	\centering
	\begin{tabular}{|l|r|r||l|r|r|}
		\hline
		\multicolumn{3}{|c||}{$v_{\mbox{max}} = \one \quad (p = \mbox{0.1})$} & \multicolumn{3}{c|}{$v_{\mbox{max}} = \five \quad (p = \mbox{0.1})$}\\
		\hline
		\hline 
		 & TCA & LWR & & TCA & LWR \\
		\hline 
		$\overline v_{\mbox{ff}}$ & 0.90 &   24.30 & $\overline v_{\mbox{ff}}$ & 4.90 &  132.30\\
		$k_{\mbox{crit}}$         & 0.50 &   66.67 & $k_{\mbox{crit}}$         & 0.17 &   22.22\\
		$k_{\mbox{jam}}$          & 0.91 &  121.20 & $k_{\mbox{jam}}$          & 0.91 &  121.20\\
		$q_{\mbox{cap}}$          & 0.45 & 1620.08 & $q_{\mbox{cap}}$          & 0.83 & 2939.71\\
		\hline
		\multicolumn{6}{c}{~}\\
		\hline
		\multicolumn{3}{|c||}{$v_{\mbox{max}} = \one \quad (p = \mbox{0.5})$} & \multicolumn{3}{c|}{$v_{\mbox{max}} = \five \quad (p = \mbox{0.5})$}\\
		\hline
		\hline 
		 & TCA & LWR & & TCA & LWR \\
		\hline 
		$\overline v_{\mbox{ff}}$ & 0.50 &  13.50 & $\overline v_{\mbox{ff}}$ & 4.50 &  121.50\\
		$k_{\mbox{crit}}$         & 0.50 &  66.67 & $k_{\mbox{crit}}$         & 0.17 &   22.22\\
		$k_{\mbox{jam}}$          & 0.67 &  88.89 & $k_{\mbox{jam}}$          & 0.67 &   88.89\\
		$q_{\mbox{cap}}$          & 0.25 & 900.05 & $q_{\mbox{cap}}$          & 0.77 & 2699.73\\
		\hline 
	\end{tabular}
	\caption{
		The resulting parameters for the triangular fundamental diagrams, as 
		calculated by means of equations \eqref{eq:LWRFDExpressionVFF} -- 
		\eqref{eq:LWRFDExpressionQCAP}. The TCA's parameters are expressed 
		in cells/time step, vehicles/cell, and vehicles/time step, respectively, 
		whereas the LWR's parameters are expressed in kilometres/hour, 
		vehicles/kilometre, and vehicles/hour, respectively.
	}
	\label{table:LWRCalculatedFDParameters}
\end{table}

		\subsection{Results and discussion}

\begin{figure}[!htb]
	\centering
	\includegraphics[width=\figurehalfwidth]{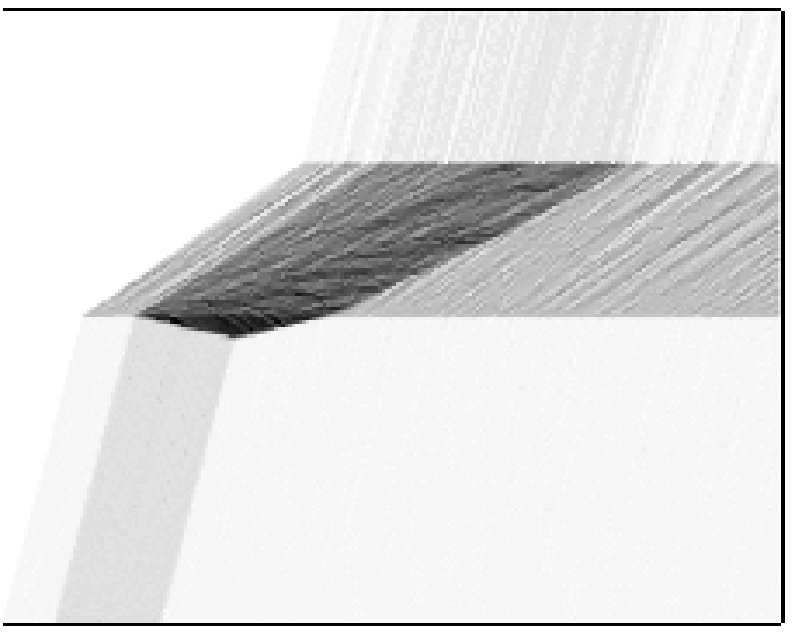}
	\hspace*{0.25cm}
	\includegraphics[width=\figurehalfwidth]{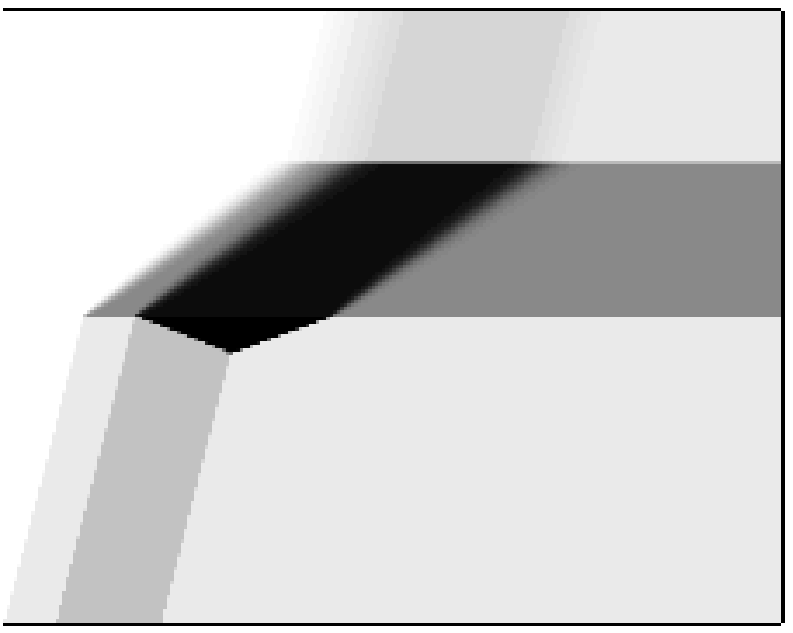}
	\caption{
		Time-space diagrams showing the propagation of densities during 3000 time 
		steps for the road in the case study. \emph{Left:} the microscopic system 
		dynamics of the STCA model. \emph{Right:} the results for the LWR model. In 
		both cases, $p =$~0.1, with darker regions corresponding to more congested 
		traffic conditions. There is a qualitatively good agreement between the two 
		approaches on the level of first-order traffic flow characteristics: the 
		buildup and dissolution of congestion queues are fairly analogous for both 
		techniques.
	}
	\label{fig:STCAAndLWRTXDensityDiagramsP01}
\end{figure}

The result of numerically solving the LWR model for the case of $p =$~0.1 using 
the Godunov method \cite{DAGANZO:95b,LEBACQUE:96}, is depicted in the right part 
of \figref{fig:STCAAndLWRTXDensityDiagramsP01}. Note that for the LWR model, 
each cell in the Godunov scheme corresponds to 5 (i.e., $v_{\mbox{max}}^{A,C}$) 
consecutive cells of the STCA model. Comparing the tempo-spatial behavior of the 
LWR model to that of the microscopic system dynamics of the STCA model (i.e., 
the left part of \figref{fig:STCAAndLWRTXDensityDiagramsP01}), we find a good 
\emph{qualitative} agreement between the two approaches. With respect to the 
first-order traffic flow characteristics, we note that the buildup and 
dissolution of congestion queues are fairly analogous for both techniques.

\begin{figure}[!htb]
	\centering
	\includegraphics[width=\figurehalfwidth]{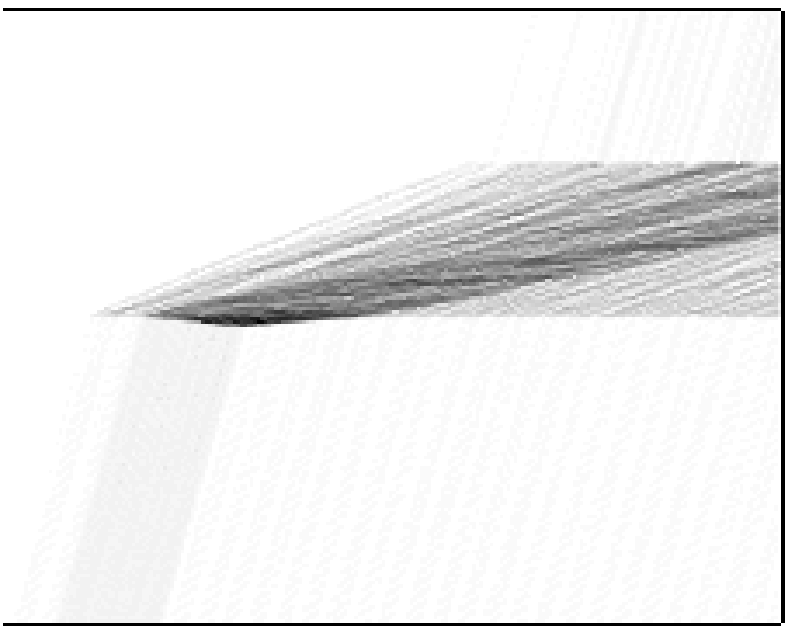}
	\hspace*{0.25cm}
	\includegraphics[width=\figurehalfwidth]{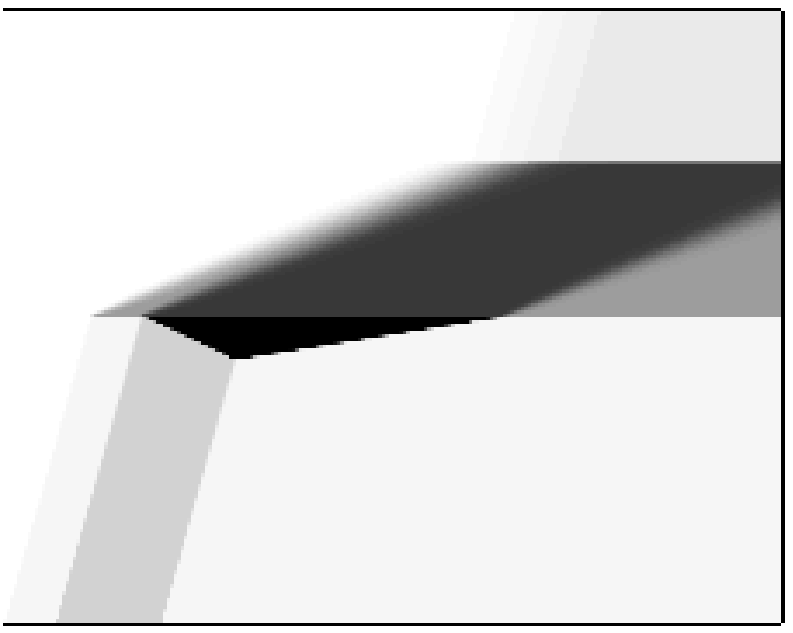}
	\caption{
		Time-space diagrams showing the propagation of densities during 3000 time 
		steps for the road in the case study. \emph{Left:} the microscopic system 
		dynamics of the STCA model. \emph{Right:} the results for the LWR model. In 
		both cases, $p =$~0.5, with darker regions corresponding to more congested 
		traffic conditions. As revealed by the shape of the dark triangular region 
		in the LWR model (right part), the buildup and dissolution of congestion 
		queues is exaggerated, especially in the upstream flowing queue of segment 
		$A$.
	}
	\label{fig:STCAAndLWRTXDensityDiagramsP05}
\end{figure}

In \figref{fig:STCAAndLWRTXDensityDiagramsP05}, we show the results when 
repeating the same experiment, but this time with the stochastic noise $p$ set 
to 0.5 for all three segments. As revealed by the shape of the dark triangular 
region in the LWR model (right part), the buildup and dissolution of congestion 
queues seems to be \emph{exaggerated}, especially in the upstream flowing queue 
of segment $A$.

It is interesting to note that the STCA model reveals a \emph{higher-order 
effect} that is not visible in the LWR model: there exists a fan of forward 
propagating density waves in segment $B$ (see the left parts of 
\figref{fig:STCAAndLWRTXDensityDiagramsP01} and 
\figref{fig:STCAAndLWRTXDensityDiagramsP05}). As such, in its tempo-spatial 
diagram, the STCA seems to be able to visualize the characteristics that 
constitute the solution of the LWR model.

\begin{figure}[!htb]
	\centering
	\includegraphics[width=\figurehalfwidth]{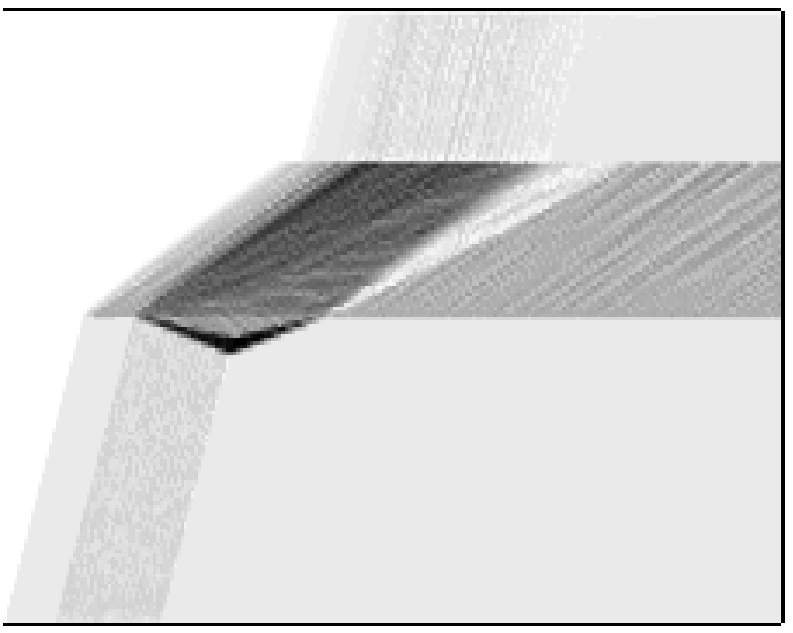}
	\hspace*{0.25cm}
	\includegraphics[width=\figurehalfwidth]{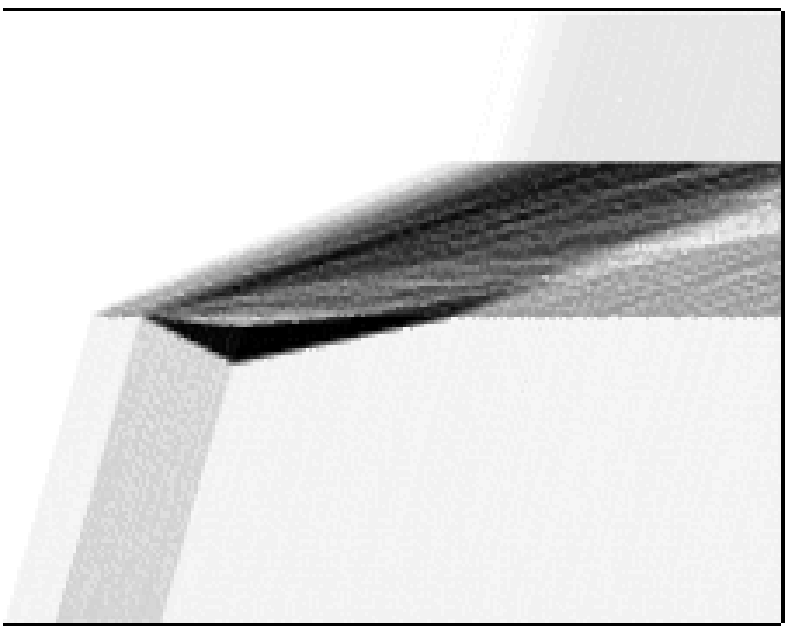}
	\caption{
		Time-space diagrams showing the differences in densities for the STCA and 
		LWR models, during 3000 time steps for the road in the case study. Darker 
		regions indicate large differences between both modeling approaches. 
		\emph{Left:} the differences for $p =$~0.1 are less pronounced, showing only 
		a dark edge at the bottom triangular-shaped region in segment $A$. 
		\emph{Right:} the differences for $p =$~0.5, showing significant 
		discrepancies in the bottom of the triangular-shaped region in segment $A$.
	}
	\label{fig:STCAAndLWRTXDensityDiagramsDifferences}
\end{figure}

In order to more rigourously quantify the discrepancies between the time-space 
diagrams of both STCA and LWR models, we provide their \emph{absolute 
differences} in \figref{fig:STCAAndLWRTXDensityDiagramsDifferences}. The left 
part shows the differences for $p =$~0.1, whereas the right part shows the 
differences for $p =$~0.5. The most important features to look at, are the dark 
colored regions which indicate larger differences between both modeling 
approaches. We can clearly see that there is a problem with respect to a 
\emph{quantitative} agreement between both STCA and LWR models. It appears as 
though the LWR model \emph{overestimates} the STCA's capacity flows. As a 
result, it dissolves its jams more quickly in segment $B$, and it predicts a 
more severe onset of congestion in segment $A$ (i.e., the triangular-shaped 
region containing the spill-back queue is more pronounced in the LWR's case). 
The sharply pronounced darker regions in the tempo-spatial left part of segment 
$B$, are due to the fact that the LWR model does not visualize the 
characteristics of its solution, in contrast to the STCA model which is able to 
give a clear indication of them.

One of the main reasons for this discrepancy between both modeling approaches, 
lies in the derivation of a triangular $q_{e}(k)$ fundamental diagram for the 
LWR model, as was explained in section \ref{sec:InitialMethodology}. Because we 
assumed a stationarity condition on the STCA's rule set, the resulting 
constraints implied an invariant critical density, and always overestimated the 
STCA's capacity flows. In our opinion, the different behavior of both models, 
mainly stems from this artifact. As a result, the discrepancies will become more 
articulated when increasing the stochastic noise $p$.

	\section{Alternate derivation of the fundamental diagram}
	\label{sec:AlternateDerivation}

Considering the results of the previous approach, i.e., deriving the LWR's 
fundamental diagram based on the STCA's rule set, and the problems related to 
it, the next step is to specify the fundamental diagram \emph{directly}, based 
on the empirically observed behavior of the STCA model. In the following two 
sections, we first discuss the effects of explicitly adding noise to the LWR's 
fundamental diagram, after which we discuss our obtained results when specifying 
the fundamental diagram directly.

		\subsection{The effect of adding noise to the LWR's fundamental diagram}

Adding noise to the LWR model can mainly be accomplished via two ways: either by 
explicitly incorporating noise terms in the LWR equations (e.g., the 
conservation equation), or as a noise term in the $q_{e}(k)$ relation (i.e., the 
fundamental diagram). We refrain from changing the LWR's conservation equation, 
because this amounts to introducing some form of numerical diffusion, similar to 
the viscosity terms that are originally encountered in the conservation 
equation's \eqref{eq:FirstOrderConservationLaw} right-hand side (which we set to 
zero in order to obtain the inviscid LWR model).

\begin{figure}[!htb]
	\centering
	\framebox{\includegraphics[width=\figurehalfwidth]{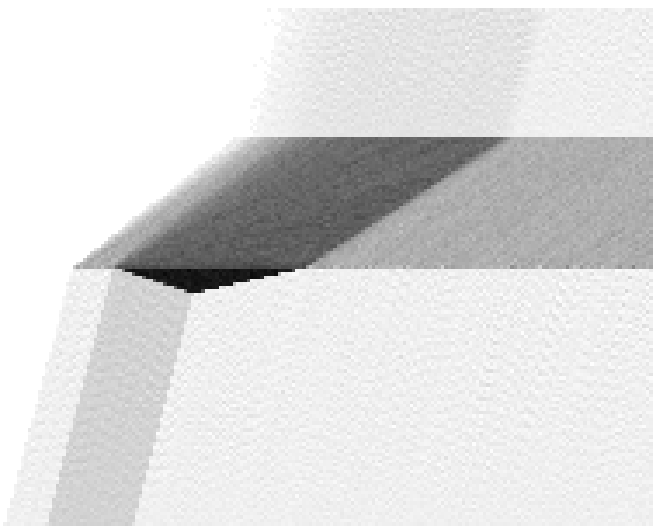}}
	\hspace*{0.25cm}
	\framebox{\includegraphics[width=\figurehalfwidth]{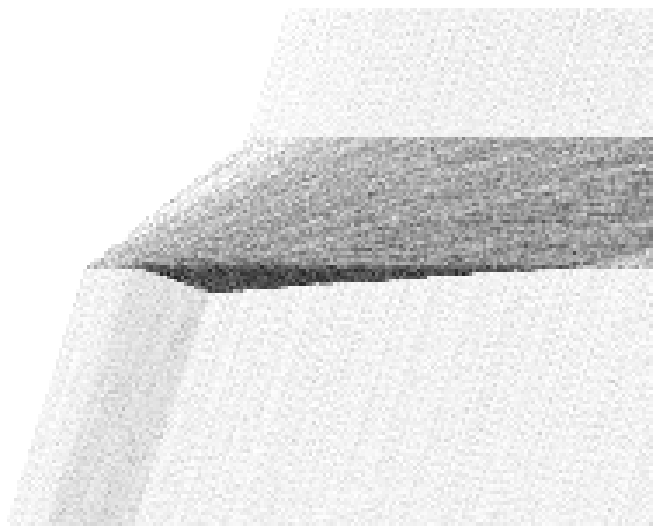}}
	\caption{
		Time-space diagrams showing the propagation of densities during 3000 time 
		steps for the road in the case study. Depicted are the results for the LWR 
		model, with noise levels of 0.1 (\emph{left}) and 0.5 (\emph{right}). Higher 
		noise levels clearly reveal the typical characteristics of the solution, and 
		introduce longer jam dissolution times.
	}
	\label{fig:LWRTXDensityDiagramsWithNoise}
\end{figure}

In \figref{fig:LWRTXDensityDiagramsWithNoise}, we show the results of supplying 
uniformly distributed additive noise of 0.1 (left part) and 0.5 (right part). As 
can be seen, the introduction of noise in the fundamental diagram, leads to a 
`spreading' of the solution. For small noise levels, some of the characteristics 
are revealed; for larger noise levels, the characteristics are clearly 
pronounced, including long jam dissolution times.

\setlength{\fboxsep}{0pt}
\setlength{\fboxsep}{\tempfboxsep}

		\subsection{Specifying the fundamental diagram directly}

\begin{figure}[!htb]
	\centering
	\includegraphics[width=\figurehalfwidth]{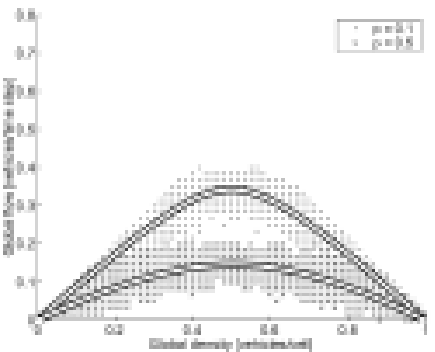}
	\hspace{0.25cm}
	\includegraphics[width=\figurehalfwidth]{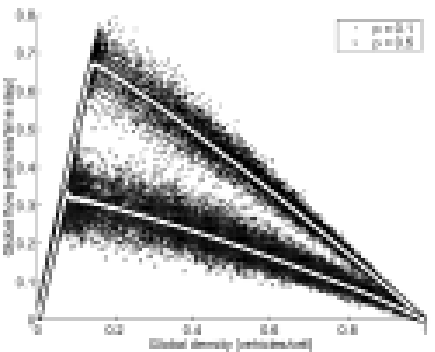}
	\caption{
		The ($k$,$q$) fundamental diagrams for the STCA model. \emph{Left:} two 
		diagrams for $v_{\mbox{max}}^{B} = \one$ cells/time step. \emph{Right:} two 
		diagrams for $v_{\mbox{max}}^{A,C} = \five$ cells/time step. Each time, the 
		slowdown probability $p \in \lbrace \mbox{0.1}, \mbox{0.5} \rbrace$. Note 
		how a slower maximum speed makes the diagrams more curved, and how an 
		increasing slowdown probability leads to both a lower critical density and 
		capacity flow.
	}
	\label{fig:STCAKQDiagrams}
\end{figure}

Instead of deriving the fundamental diagram based on the approach taken in 
section \ref{sec:InitialMethodology}, we now try to obtain the values for the 
critical densities and capacity flows directly, by looking at the STCA's 
($k$,$q$) diagrams in \figref{fig:STCAKQDiagrams} (an equivalent procedure would 
be to measure the capacity flows directly from the time-space diagram in 
\figref{fig:TCATrajectoriesCloseup}). Considering the STCA's ($k$,$q$) diagrams 
in \figref{fig:STCAKQDiagrams}, we can estimate its capacities at approximately 
$q_{\mbox{cap}}^{B} = $~0.34 vehicles/time step~$\approx$~1220 vehicles/hour, 
and $q_{\mbox{cap}}^{A,C} = $~0.67 vehicles/time step~$\approx$~2400 
vehicles/hour for $v_{\mbox{max}}^{B} = \one$ and $v_{\mbox{max}}^{A,C} = \five$ 
cells/time step, respectively. The stochastic noise $p$ was set to 0.1 for all 
three segments. Changing $p$ to 0.5 for these segments, we can estimate the 
capacities at approximately $q_{\mbox{cap}}^{B} =$~0.15 vehicles/time 
step~$\approx$~540 vehicles/hour, and $q_{\mbox{cap}}^{A,C} = $~0.34 
vehicles/time step~$\approx$~1220 vehicles/hour for $v_{\mbox{max}} = \one$ and 
$v_{\mbox{max}} = \five$ cells/time step, respectively.

\begin{figure}[!htb]
	\centering
	\includegraphics[width=\figurehalfwidth]{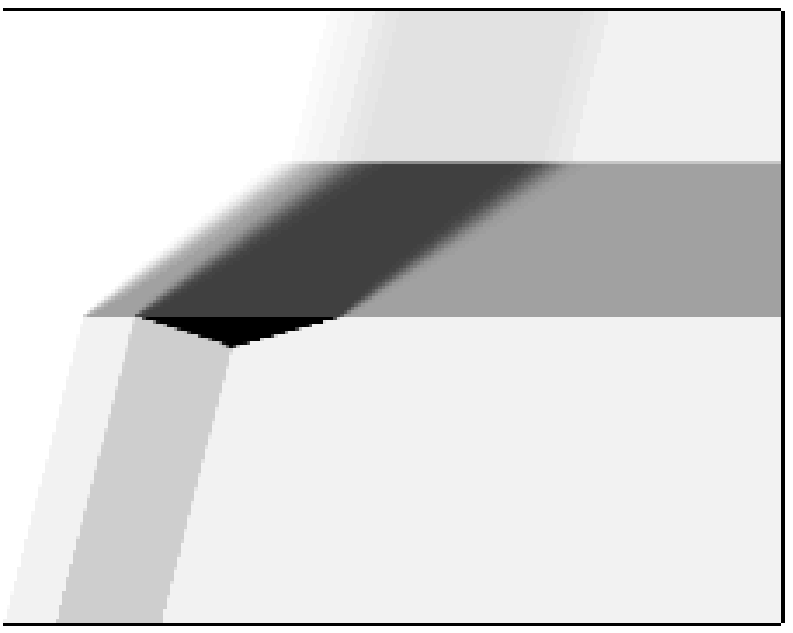}
	\hspace*{0.25cm}
	\includegraphics[width=\figurehalfwidth]{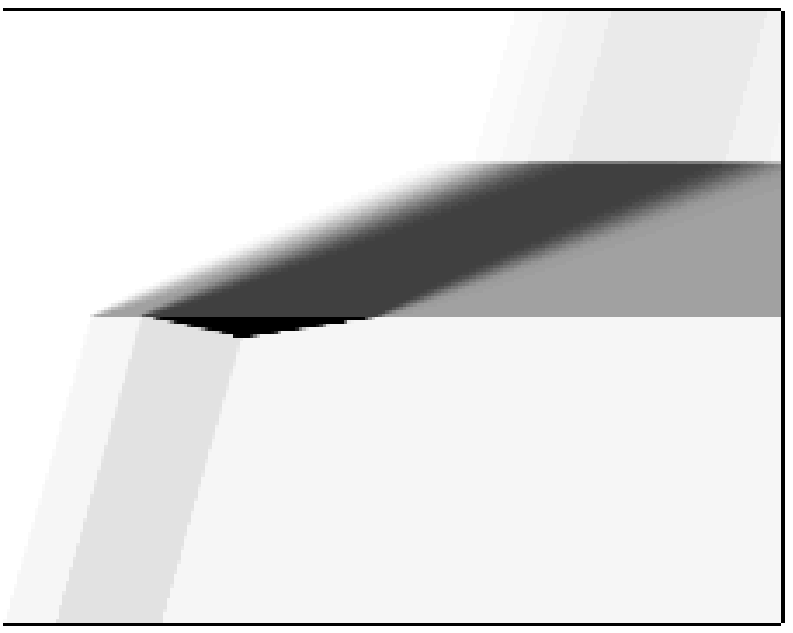}
	\caption{
		Time-space diagrams showing the propagation of densities during 3000 time 
		steps for the road in the case study. \emph{Left:} the results for the LWR 
		model with $p =$~0.1. \emph{Right:} the results for the LWR model with $p 
		=$~0.5. In both diagrams, the fundamental diagram was specified directly to 
		the LWR model, by explicitly stipulating the capacity flows of the STCA 
		model. As a result, there seems to be a better qualitative agreement for 
		both noise levels with the STCA's time-space diagrams in the left parts of 
		\figref{fig:STCAAndLWRTXDensityDiagramsP01} and 
		\figref{fig:STCAAndLWRTXDensityDiagramsP05}.
	}
	\label{fig:LWRTXDensityDiagramsP01P05}
\end{figure}

Instead of calculating the capacity flows from the average free-flow speeds and 
the critical densities, as was done by means of equation 
\eqref{eq:LWRFDExpressionQCAP}, we now specify these capacity flows directly to 
the LWR's fundamental diagrams and calculate the critical densities from them. 
The results we obtained, are visualized in the time-space diagrams of 
\figref{fig:LWRTXDensityDiagramsP01P05}. Because the STCA's capacity flows are 
now better approximated, and not overestimated as with the previous methodology, 
there seems to be a better qualitative agreement for both noise levels with the 
STCA's time-space diagrams in the left parts of 
\figref{fig:STCAAndLWRTXDensityDiagramsP01} and 
\figref{fig:STCAAndLWRTXDensityDiagramsP05}.

\begin{figure}[!htb]
	\centering
	\includegraphics[width=\figurehalfwidth]{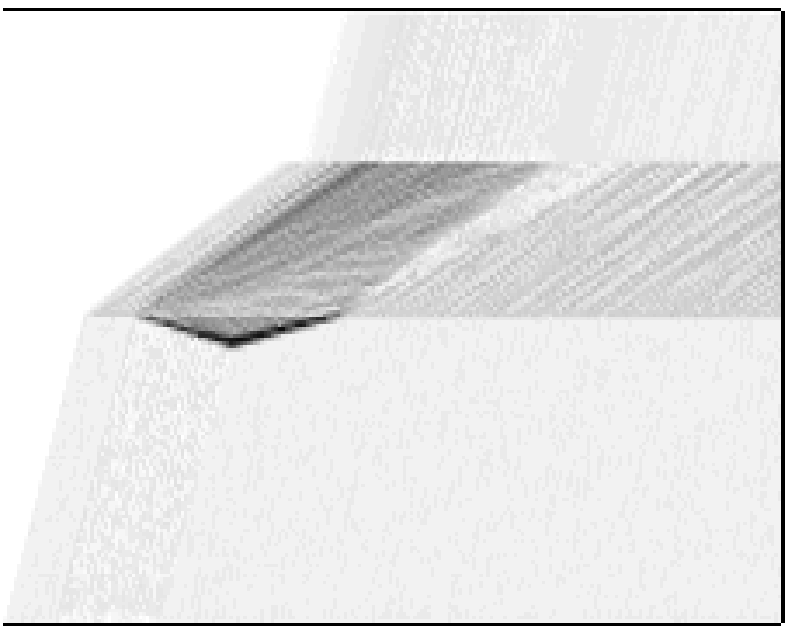}
	\hspace*{0.25cm}
	\includegraphics[width=\figurehalfwidth]{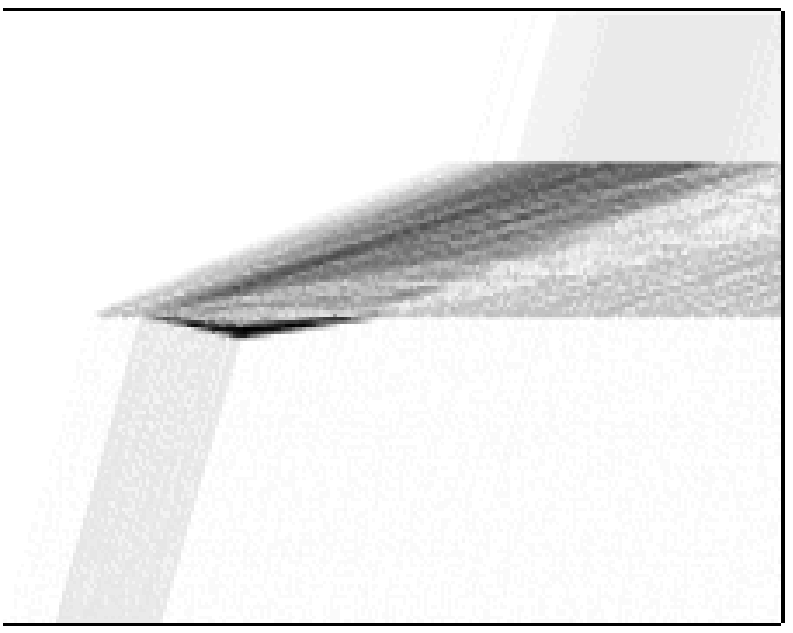}
	\caption{
		Time-space diagrams showing the differences in densities for the STCA and 
		LWR models, during 3000 time steps for the road in the case study. Darker 
		regions indicate large differences between both modeling approaches. 
		\emph{Left:} the differences for $p =$~0.1. \emph{Right:} the differences 
		for $p =$~0.5. In both cases, the differences are less pronounced, showing 
		only dark edges at the bottom of the triangular-shaped region in segment 
		$A$.
	}
	\label{fig:STCAAndLWRTXDensityDiagramsDifferencesDirect}
\end{figure}

In \figref{fig:STCAAndLWRTXDensityDiagramsDifferencesDirect} we have depicted 
the absolute differences between this approach and the STCA's time-space 
diagrams. Comparing this to the previous results of 
\figref{fig:STCAAndLWRTXDensityDiagramsDifferences}, we can see that in both 
cases the buildup and dissolution of congestion queues is in good qualitative 
agreement for both noise levels. \emph{As such, we come the conclusion that it 
is vital to correctly capture the capacity flows of the STCA model.} Neglecting 
this property, can result in severe distortion of the system dynamics for higher 
noise levels.

Note that the LWR model is able to correctly capture the first-order effects of 
jam buildup and dissolution, and that, due to its microscopic treatment, the 
STCA model allows us to visualize the higher-order effects inside jam. However, 
as is evidenced by this and the previous section, it is very important to 
correctly capture the capacity flows in the STCA model, otherwise a growing 
discrepancy between the LWR and STCA model is introduced with higher noise 
levels.

	\section{Conclusions}
	\label{sec:Conclusions}

In this paper, we presented an alternate methodology for implicitly 
incorporating the STCA's stochasticity into the macroscopic first-order LWR 
model. The innovative aspect of our approach, is that we derive the LWR's 
fundamental diagram directly from the STCA's rule set, by assuming a 
stationarity condition that converts the STCA's rules into a set of linear 
inequalities. In turn, these constraints define the shape of the fundamental 
diagram that is then specified to the LWR model.

For noise-free systems, our method is exact. In the presence of noise, however, 
the capacity flows in the derived fundamental diagram are overestimates of those 
of the STCA model. This discrepancy can be explained as follows: the underlying 
assumption for the LWR model, is that the fundamental diagram is assumed to be 
exact, and implicitly obeyed, i.e., the existing equilibrium relation is 
representative for the real traffic situation. In the original LWR formulation, 
this relation was also assumed to hold also for non-stationary traffic (which is 
a more or less reasonable assumption if we consider long and crowded roads). Our 
calculations have shown that a direct translation of the STCA's rule set into 
the LWR's fundamental diagram, does not always result in a valid fundamental 
diagram, especially for higher noise levels. As such, there can be a significant 
difference between an \emph{average} fundamental diagram (STCA) and a 
\emph{stationary} fundamental diagram (LWR). As a result, the STCA model is able 
to temporarily operate under larger flows and densities than those possible for 
the LWR's stationary fundamental diagram. A logical course of action would to 
better approximate the STCA's fundamental diagram. By doing so however, we lose 
the advantage gained through an explicit derivation of the fundamental diagrams' 
outer envelope, almost certainly leading to extra conditions that need to make 
further assumptions about its shape. Directly specifying the STCA's capacity 
flows to the triangular LWR fundamental diagram, effectively remedies most of 
the mismatches between both STCA and LWR models.

Our methodology sees the STCA complementary to the LWR model and vice versa, so 
the results can be of great assistance when interpreting the traffic dynamics in 
both models. Especially appealing, is the fact that the STCA can visualise the 
higher-order characteristics of traffic stream dynamics, e.g., the fans of 
rarefaction waves. Nevertheless, because the LWR model is only a coarse 
representation of reality, there are still some mismatches between the two 
approaches. One of the main concerns the authors discovered, is as hinted at 
earlier, the fact that using a stationary fundamental diagram (i.e., an 
equilibrium relation between density and flow), always overestimates the 
practical capacity of a stochastic cellular automaton model. As such, it is 
vital to correctly capture the capacity flows in both STCA and LWR models, a 
remark that we feel is valid for all case studies.


\section*{Acknowledgements}

\begin{acknowledgement}
	Dr. Bart De Moor is a full professor at the Katholieke Universiteit Leuven, Belgium.
  \noindent
  Our research is supported by:
  \textbf{Research Council KUL}: GOA AMBioRICS, several PhD/post\-doc
  \& fellow grants,
	\textbf{Flemish Government}:
  \textbf{FWO}: PhD/post\-doc grants, projects, G.0407.02 (support vector machines),
  G.\-0197.02 (power islands), G.0141.03 (identification and cryptography), G.0491.03
  (control for intensive care glycemia), G.0120.\-03 (QIT), G.0452.04 (new quantum algorithms),
  G.0499.04 (statistics), G.0211.05 (Nonlinear), research communities (ICCoS, ANMMM, MLDM),
  \textbf{IWT}: PhD Grants, GBOU (McKnow),
  \textbf{Belgian Federal Science Policy Office}: IUAP P5/\-22 (`Dynamical Systems and
  Control: Computation, Identification and Modelling', 2002-2006), PODO-II (CP/40:
  TMS and Sustainability),
  \textbf{EU}: FP5-Quprodis, ERNSI,
  \textbf{Contract Research/agreements}: ISMC/IPCOS, Data4s,TML, Elia, LMS,
  Mastercard.
\end{acknowledgement}

\bibliography{paper}

\end{document}